\documentclass[11pt,a4paper]{article}
\pdfoutput=1
\usepackage{graphicx,epsfig}
\usepackage{dcolumn} % Align table columns on decimal point
\usepackage{slashed,color,amsmath,amssymb}
\usepackage{jheppub}
\usepackage{enumitem}
\usepackage{mathrsfs}

\usepackage{caption, subcaption}
\usepackage{multirow}
\usepackage{verbatim}
\newcommand{\be}{\begin{equation}}
\newcommand{\ee}{\end{equation}}
\newcommand{\bea}{\begin{eqnarray}}
\newcommand{\eea}{\end{eqnarray}}
\newcommand{\bit}{\begin{itemize}}
\newcommand{\eit}{\end{itemize}}

\def\gsim{\lower0.5ex\hbox{$\:\buildrel >\over\sim\:$}}
\def\lsim{\lower0.5ex\hbox{$\:\buildrel <\over\sim\:$}}

\hypersetup{
   colorlinks=true,       % false: boxed links; true: colored links
   linkcolor=blue,        % color of internal links
   citecolor=red,         % color of links to bibliography
   filecolor=magenta,     % color of file links
}

\usepackage{bm}

\preprint{}
\title{Exploring $Z'$ and Right-Handed Neutrinos in the BLSM at the Large Hadron Collider}

\author{Nidal Chamoun$^{1,2,5}$, Kareem Ezzat$^{3,4}$, Shaaban Khalil$^4$, Rhitaja Sengupta$^5$}

\affiliation{\vspace*{0.1in}$^1$ Department of Statistics, Faculty of Science, Damascus University, Syria.}
\affiliation{\vspace*{0.1in}$^2$ CASP, Antioch Syrian University, Maaret Saidnaya, Damascus, Syria.}
\affiliation{\vspace*{0.1in}$^3$ Department of Mathematics, Faculty of Science, Ain Shams University, Cairo 11566, Egypt.}
\affiliation{\vspace*{0.1in}$^4$ Center for Fundamental Physics, Zewail City of Science and
	Technology, 6th of October City, Giza 12578, Egypt}
\affiliation{\vspace*{0.1in}$^5$ Bethe Center for Theoretical Physics and Physikalisches Institut der Universit\"at Bonn, \\
	Nußallee 12, 53115 Bonn, Germany}

\emailAdd{chamoun@uni-bonn.de}
\emailAdd{kareemezat@sci.asu.edu.eg}
\emailAdd{skhalil@zewailcity.edu.eg}
\emailAdd{rsengupt@uni-bonn.de}

\preprint{BONN-TH-2024-20}

\abstract{
We study the collider phenomenology of the $B$-$L$ extension of the Standard Model (BLSM), focusing on the production and decay of a heavy neutral gauge boson (\( Z' \)) at the Large Hadron Collider (LHC). In this framework, the \( Z' \) can decay into pairs of heavy right-handed neutrinos (\( \nu_R \)), which subsequently decay into charged leptons and \( W \) bosons. These processes give rise to three distinctive final states: (i) two leptons plus four jets (\( 2\ell + 4j \)), (ii) four leptons plus missing transverse energy (\( 4\ell + \text{MET} \)), and (iii) three leptons plus two jets and MET (\( 3\ell + 2j + \text{MET} \)).
To enhance signal sensitivity and suppress Standard Model backgrounds, we employ multivariate analysis techniques based on Boosted Decision Trees (BDTs), as well as selection optimizations using the \texttt{XGBOOST} framework. The classifiers are trained on kinematic observables sensitive to the masses of the \( Z' \) and \( \nu_R \). We demonstrate that all three final states offer significant discovery potential for both the \( Z' \) and heavy \( \nu_R \) at the High-Luminosity LHC. Our results highlight the testability of the BLSM at current and future collider experiments, and provide a promising avenue for probing the origin of neutrino masses and the baryon asymmetry of the Universe.
}

\begin{document}

\maketitle

%\tableofcontents

%%%%%%%%%%%%%%%%%%%%%%%%%%%%%%%%%%

\section{Introduction}
\label{sec:intro}

The Standard Model (SM) of particle physics has been highly successful in explaining a wide range of phenomena, but it falls short in addressing certain critical issues. Notably, it cannot explain neutrino oscillations, which indicate that neutrinos have mass \cite{Fukuda:1998mi,Ahn:2002up,Eguchi:2002dm,Ahmad:2002jz,An:2012eh}, nor the observed baryon asymmetry in the universe, where matter significantly outweighs antimatter. These deficiencies suggest that an extension of the SM is required to account for these phenomena.

One promising extension involves the introduction of right-handed neutrinos (RHN), which can naturally explain the small but non-zero masses of neutrinos through the seesaw mechanism. Additionally, extending the gauge symmetry of the SM provides a framework to incorporate new physics without fundamentally altering its successful aspects. Specifically, the $B$-$L$ (baryon minus lepton number) extension of the SM (BLSM)\cite{Mohapatra:1980qe,Papaefstathiou:2011rc,Masiero:1982fi,Mohapatra:1982xz,Buchmuller:1991ce} offers a minimal and elegant approach to address these shortcomings.

The minimal extension is achieved by augmenting the SM gauge group with an additional \( U(1)_{B\mbox{-}L} \) symmetry, leading to the gauge group:
\begin{equation}
G_{B\mbox{-}L} = SU(3)_C \times SU(2)_L \times U(1)_Y \times U(1)_{B\mbox{-}L}.
\end{equation}
The BLSM extension introduces new particles and interactions that contribute to the SM phenomenology, particularly in the neutrino and baryon sectors.
\begin{enumerate}
\item Right-handed Neutrinos:   The model introduces three RHN \( \nu^i_R \) (where \( i = 1,2,3 \)), each with a $B$-$L$ charge of \(-1\). These particles do not interact with the SM weak forces but are crucial for generating neutrino masses via the seesaw mechanism.

\item New Gauge Boson \( Z' \):  The extension includes an extra gauge boson, denoted as \( Z' \), corresponding to the $B$-$L$ gauge symmetry. This new force carrier could have observable signatures at high-energy colliders such as the LHC, providing a direct test of the extended model.

\item Additional Scalar Field \( \chi \):  A new scalar field \( \chi \), which is a singlet under the SM gauge group but carries a $B$-$L$ charge of \( +2 \), is introduced. This scalar plays a crucial role in breaking the \( U(1)_{B\mbox{-}L} \) symmetry, analogous to how the Higgs field breaks the electroweak symmetry in the SM.
\end{enumerate}

In this paper, we investigate the experimental signatures of these new particles, particularly the  RHN and the \( Z' \) boson, at the Large Hadron Collider (LHC). This question has been addressed previously using different methods and processes, as discussed in \cite{Kang:2015uoc,Das:2022rbl}. However, we present in this work a distinct computational approach, based on machine learning techniques (XGBOOST, BDT) and consider different production processes, in such a way to
justify a fresh investigation of these signatures, not only updating the constraints but also exploring the discovering potential even within a well-studied theoretical
framework. The \( Z' \) boson, for instance, can manifest as a resonance in dilepton or dijet final states, serving as a potential smoking gun for physics beyond the SM. Similarly, RHNs open up novel decay channels and distinct collider signatures, offering valuable insights into the origin of neutrino masses and the baryon asymmetry of the universe.  We emphasize that these signatures not only probe the existence of these particles but also serve as critical tests of the BLSM, which seeks to address several fundamental questions in particle physics. Specifically, we examine four distinct signals arising from \( Z' \) production at the LHC, where the \( Z' \) decays into two RHNs, (\( \nu_R \)). These decay channels lead to the following final states:  
\begin{enumerate}[label=~\roman*.]
\item Two leptons (same-sign or opposite-sign), four jets. 
\item Four leptons and missing energy.  
\item Three leptons, two jets, and missing energy.   
\item  One lepton, multiple jets, and missing energy.  
\end{enumerate}
The first three channels arise from the decay of two RHNs into $W$ boson and lepton, with the $W$ boson subsequently decaying either into lepton and $\nu_L$ or into quarks. The fourth channel occurs when one RHN decays into a $W$ boson and a lepton, with the $W$ decaying into jets, while the second RHN decays into a $Z$ boson and $\nu_L$, and the $Z$ decays into two $\nu_L$ or two jets. 

The decay width of the RHN into a \( Z \) boson and \( \nu_L \) is significantly smaller than its decay into a \( W \) boson and a charged lepton. As a result, the production cross section for the fourth channel is substantially lower compared to the other channels. Moreover, this channel suffers from considerable SM background. Therefore, our analysis in the subsequent sections will focus exclusively on the first three channels.
We employ Boosted Decision Tree (BDT) techniques~\cite{Xia:2018cfz,XIA201915,quinlan1986induction,quinlan1987simplifying} to optimize the selection criteria and enhance the sensitivity to the \( Z' \) and RHN signals, ensuring effective discrimination between potential new physics signatures and SM background processes. In addition, we utilize the \texttt{XGBOOST} algorithm (eXtreme Gradient Boosting)~\cite{Xia:2018cfz,XIA201915,quinlan1986induction,quinlan1987simplifying} to further refine the event selection and to identify the most discriminating kinematic variables that separate signal from background.

The structure of the paper is as follows: In Section~2, we present a concise review of the BLSM, with emphasis on the \( Z' \) boson and the RHN. We discuss the origin of their masses from the spontaneous breaking of the \( B\mbox{-}L \) symmetry and summarize the key interactions relevant to our analysis. This section establishes the theoretical framework for the collider signatures explored in the subsequent sections.  Section~3 outlines the computational methodology, including the application of BDT techniques and the implementation of the \texttt{XGBOOST} algorithm to optimize signal-background discrimination.  In Section~4, we analyze the four distinct final-state signatures arising from \( Z' \) production at the LHC, focusing on their kinematic features and discovery potential.  Finally, Section~5 presents our conclusions and discusses future directions.

%%%%%%%%%%%%%%%%%%%%%%%%%%%%%%%%%%%%%%%%%%%%
\section{$Z'$ and $\nu_R$ in the BLSM}
\label{sec:model}

As discussed above, the minimal BLSM is based on the gauge group $SU(3)_C \times SU(2)_L \times U(1)_Y \times U(1)_{B-L}$. The Lagrangian for this $B\mbox{-}L$ extension is expressed as follows:
\begin{eqnarray}
{\cal L}_{B-L}
&=&
i{\bar l}D_{\mu}\gamma^{\mu}l+i{\bar e_R}D_{\mu}\gamma^{\mu}e_R+i{\bar \nu_R}D_{\mu}\gamma^{\mu}\nu_R
+
i{\bar Q}D_{\mu}\gamma^{\mu}Q+i{\bar d_R}D_{\mu}\gamma^{\mu}d_R+i{\bar u_R}D_{\mu}\gamma^{\mu}u_R
\nonumber \\
&-&
\frac14G^\alpha_{\mu\nu}G^{\mu\nu\alpha}-\frac14W_{\mu\nu}W^{\mu\nu}-\frac14B_{\mu\nu}B^{\mu\nu}-\frac14C_{\mu\nu}C^{\mu\nu}
+ (D^{\mu}\phi)^{\dagger}(D_{\mu}\phi)+(D^{\mu}\chi)^{\dagger}(D_{\mu}\chi)
\nonumber\\
&-&
m^2_1\phi^{\dagger}\phi-m^2_2\chi^{\dagger}\chi-\lambda_1(\phi^{\dagger}\phi)^2-\lambda_2(\chi^{\dagger}\chi)^2-\lambda_3(\phi^{\dagger}\phi)(\chi^{\dagger}\chi)
- \lambda_{e}{\bar l}\phi e_R-\lambda_{\nu}{\bar l}{\tilde \phi}\nu_R \nonumber \\
&-& \frac12\lambda_{\nu_R}{\bar\nu}^c_R\chi\nu_R-h.c. - \lambda_{d}{\bar Q}\phi d_R-\lambda_{u}{\bar Q}{\tilde \phi}u_R-h.c.,
\end{eqnarray}
where $C_{\mu\nu} = \partial_{\mu} C_{\nu} - \partial_{\nu} C_{\mu}$ represents the field strength tensor of the $U(1)_{B\mbox{-}L}$ gauge field. The covariant derivative $D_{\mu}$ is given by:
\begin{equation}\label{Gequ}
D_{\mu} =
\partial_{\mu}-ig_sT^\alpha G^\alpha_\mu-ig_2A^a_{\mu}\tau^a-ig'YB_{\mu}-i(\tilde{g}Y+g'' Y_{B-L})C_{\mu},~~~
\tau^a=\frac{\sigma^a}{2},~{\rm and}~ \sigma^a {\rm are ~Pauli~matrices}. 
\end{equation}
Here, $g''$, called also $g_{_{B\mbox{-}L}}$, is the coupling constant of the $U(1)_{B\mbox{-}L}$ gauge group, and $Y_{B\mbox{-}L}$ denotes the $B\mbox{-}L$ quantum numbers of the particles involved, which are listed in Table \ref{tabel1}.
\begin{table}[thb]
	\begin{center}
		\resizebox{\textwidth}{!}{ 
			\begin{tabular}{|c|cccccccc|} \hline
				& $l$ & $\nu_R$ & $e_R $ & $Q$ & $u_R$ & $d_R$ &$\phi$ & $\chi$ 
				\\ \hline
				$SU(3)_C\times SU(2)_L\times U(1)_Y$ 
				& $(1,{\bf 2}, -1/2)$  &  $(1,{\bf 1}, 0)$  &  $(1,{\bf 1}, -1)$
				& $(3,{\bf 2}, 1/6)$&  $(3,{\bf 1}, 2/3)$
				&  $(3,{\bf 1}, -1/3)$ & $(1,{\bf 2}, 1/2)$  &  $(1,{\bf 1}, 0)$  
				\\ \hline
				$U(1)_{B\mbox{-}L}$ & $-1$  &  $-1$  &  $-1$
				& $1/3$&  $1/3$&  $1/3$ & 0& 2
				\\ \hline
		\end{tabular}}
		\caption{The $SU(3)_C\times SU(2)_L \times U(1)_Y$ and $U(1)_{B\mbox{-}L}$ quantum numbers of the particles in the model.}
		\label{tabel1}
	\end{center}
\end{table}

To analyze the breaking of $B\mbox{-}L$ and electroweak symmetries, we consider the most general Higgs potential that remains invariant under these symmetries. The potential is expressed as:
\begin{equation}
V(\phi,\chi) = m_1^2 \phi^\dagger \phi + m_2^2 \chi^\dagger \chi + \lambda_1 (\phi^\dagger \phi)^2 + \lambda_2 (\chi^\dagger \chi)^2 + \lambda_3 (\chi^\dagger \chi)(\phi^\dagger \phi),
\end{equation}
where $\lambda_3 > -2\sqrt{\lambda_1 \lambda_2}$ and $\lambda_1, \lambda_2 \geq 0$ to ensure that the potential is bounded from below. These conditions represent the stability requirements of the potential.
Furthermore, to prevent $\langle \phi \rangle = \langle \chi \rangle = 0$ from being a local minimum, we assume that $\lambda_3^2 < 4 \lambda_1 \lambda_2$. Similar to the usual Higgs mechanism in the SM, the vacuum expectation values (vevs) $v$ for $\phi$ and $v'$ for $\chi$ can only emerge if the squared mass terms are negative, i.e., $m_1^2 < 0$ and $m_2^2 < 0$.

It is worth noting that the most general kinetic Lagrangian of the BLSM allows for gauge kinetic mixing between \( U(1)_Y \) and \( U(1)_{B\mbox{-}L} \). This mixing can be absorbed through a redefinition of the covariant derivative where, before the transformation of the gauge coupling matrix, the covariant derivative can be defined as:
\begin{equation}
	i(D_{\mu}-\partial_{\mu})=g_{YY}YA_\mu^1+g_{YB}YA_\mu^2+g_{BY}Y_{B\mbox{-}L}A_\mu^1+g_{BB}Y_{B\mbox{-}L}A_\mu^2
\end{equation}
where $A_\mu^1$ and $A_\mu^2$ are gauge fields for \( U(1)_Y \) and \( U(1)_{B\mbox{-}L} \) respectively, so we can use the next transformation to absorb the mixing between  \( U(1)_Y \) and \( U(1)_{B\mbox{-}L} \), and get Eq. (\ref{Gequ}):
\begin{eqnarray}
	G = \begin{pmatrix}
		g_{_{YY}} & g_{_{YB}} \\
		g_{_{BY}} & g_{_{BB}} \\
	\end{pmatrix}
	~~ \Longrightarrow ~~
	\tilde{G} = \begin{pmatrix}
		g' & \tilde{g} \\
		0 & g'' \\
	\end{pmatrix},
\end{eqnarray}
where 
\begin{eqnarray}
	g' = \frac{g_{_{YY}} g_{_{BB}} - g_{_{YB}} g_{_{BY}}}{\sqrt{g_{_{BB}}^2 + g_{_{BY}}^2}}, \quad 
	g'' \equiv g_{_{B\mbox{-}L}} = \sqrt{g_{_{BB}}^2 + g_{_{BY}}^2}, \quad 
	\tilde{g} = \frac{g_{_{YB}} g_{_{BB}} + g_{_{BY}} g_{_{YY}}}{\sqrt{g_{_{BB}}^2 + g_{_{BY}}^2}}.
\end{eqnarray}
Here, $g'$  is the \( U(1)_Y \) gauge coupling, $g_2$ the gauge coupling of \( SU(2)_L \) and $g_s$ the gauge coupling of \( SU(3)_C \), and we assume gauge coupling unification  at the GUT scale. 

In this basis, after $B$-$L$ and electroweak symmetry breaking, the masses of the \( Z \) and \( Z' \) bosons are given by: 
\begin{eqnarray}\label{mz1}
	M_Z^2 = \frac{1}{4} (g'^2 + g_2^2) v^2, \quad 
	M_{Z'}^2 = 4g''^2 v'^2 + \frac{1}{4} \tilde{g}^2 v^2,
\end{eqnarray}

Furthermore, the mixing angle ($\theta'$) between the $Z$ and $Z'$ bosons is given by: 
\begin{eqnarray}
	\tan 2\theta' = \frac{2 \tilde{g} \sqrt{g'^2 + g_2^2}}{\tilde{g}^2 + 16 \left( \frac{v'}{v} \right)^2 g''^2 - g_2^2 - g'^2}.
\end{eqnarray}

%The experimental searches at high energies place lower bounds on the mass of this extra neutral gauge boson. The most stringent constraint comes from the LEP II experiment~\cite{Carena:2004xs}. As an $e^+ e^-$ collider, LEP II was highly effective in constraining additional gauge bosons that significantly couple to electrons. Moreover, recent results from the CDF II experiment~\cite{Basso:2009hf,Chun:2017spz} are consistent with the LEP II constraints on the $Z^{\prime}$ mass in the context of the $B-L$ extension of the SM. Based on these findings, the typical lower bound on $M_{Z^\prime}$ is:
%\be
%\frac{M_{Z^{\prime}}}{g''} > 7~\text{TeV}.
%\ee

From Eq.~(\ref{mz1}), this implies that $v' \gtrsim \mathcal{O}(\text{TeV})$, where $v$ of order SM vev. In models with an extra Abelian gauge symmetry \( U(1)_{B\mbox{-}L} \) and kinetic mixing between the hypercharge gauge field \( B_\mu \) and the new gauge boson \( B'_\mu \), the physical coupling of the \( Z' \) boson to SM fermions is modified.
The effective coupling of \( Z' \) to a fermion \( f \) is given by ~\cite{Basso:2009hf,Chun:2017spz}:
\begin{equation}
	g_f^{Z'} = g_{B\mbox{-}L} \cdot Q_{B\mbox{-}L}^f + \tilde{g} \cdot Y^f,
\end{equation}
For the benchmark point in Table (\ref{tab:benchmark})
and considering charged leptons (\( e, \mu \)) with charges:
\[
Q_{B-L} = -1, \quad Y = -1,
\]
the effective coupling becomes:
\begin{align}
	g_{\ell}^{Z'} &= 0.42 \cdot (-1) + (-0.55) \cdot (-1) \\
	&= -0.42 + 0.55 = 0.13.
\end{align}
The cross section for the process \( pp \to Z' \to \ell^+ \ell^- \) is approximately proportional to:
\begin{equation}
	\sigma(pp \to Z') \times \text{BR}(Z' \to \ell^+ \ell^-) \propto \left( \sum_{q} |g_q^{Z'}|^2 \right) \cdot \frac{|g_\ell^{Z'}|^2}{\sum_f |g_f^{Z'}|^2}.
\end{equation}
Thus, a smaller \( g_\ell^{Z'} \) leads to a suppressed dilepton signal. For \( g_\ell^{Z'} = 0.13 \), the current LHC bounds (e.g., from ATLAS and CMS Run 2) allow:
\begin{equation}
M_{Z'} \gtrsim 2.5 - 3~\text{TeV},
\end{equation}

In addition, after the $U(1)_{B\mbox{-}L}$ symmetry breaking~\cite{Khalil:2006yi}, the Yukawa interaction term $\lambda_{\nu_R}\chi \bar{\nu}^c_{R}\nu_{R}$ generates a mass for the RHN, given by:
\begin{equation}
M_R = \frac{1}{2\sqrt{2}}\lambda_{\nu_R} v'.
\end{equation}
Moreover, electroweak symmetry breaking results in the Dirac neutrino mass term:
\begin{equation}
m_D = \frac{1}{\sqrt{2}}\lambda_\nu v.
\end{equation}
Thus, the mass matrix for the left- and right-handed neutrinos is:
\be
\left( \begin{array}{cc}
	0 & m_D \\
	m_D & M_R \\
\end{array} \right).
\ee
Since \( M_R \) is proportional to \( v' \) and \( m_D \) is proportional to \( v \), it follows that \( M_R > m_D \). Diagonalizing this mass matrix yields the following masses for the light ($\nu_L$) and heavy ($\nu_H$) neutrinos, respectively:
\begin{equation}
m_{\nu_L} = -m_D M_R^{-1} m_D^T,
\ee
\be
m_{\nu_H} = M_R.
\end{equation}
Thus, the $B\mbox{-}L$ gauge symmetry provides a natural framework for the seesaw mechanism. However, the scale of $B\mbox{-}L$ symmetry breaking, $v'$, remains arbitrary. As in \cite{Khalil:2006yi}, $v'$ is assumed to be of order TeV, making $M_R$ also of that scale.

The $Z'$ boson can be produced in hadron colliders, such as the LHC, through its couplings to quarks. While it can decay into various final states \cite{atlas2019zprime,leike1999zprime,langacker2009zprime}, we focus on the scenario where the $Z'$ decays into a pair of RHNs. This scenario is particularly interesting because the subsequent decay of the neutrinos into a lepton and a $W$ boson can lead to a final state with two same-sign leptons, when the $W$ boson decays into jets. 

The cross-section for the process \( pp \to Z' \to \nu_R \nu_R \) can be expressed as:
\begin{equation}
\sigma(pp \to Z' \to \nu_R \nu_R) = \int dx_1 dx_2 \, f_{q}(x_1, Q^2) f_{\bar{q}}(x_2, Q^2) \, \hat{\sigma}(\hat{s}),
\end{equation}
where $f_{q}(x_1, Q^2)$ and $f_{\bar{q}}(x_2, Q^2)$ are the parton distribution functions (PDFs) for the quark and antiquark inside the protons \cite{cteq1995pdfs,nnpdf2015pdfs}, evaluated at momentum fractions $ x_1 $ and $x_2 $, and scale $ Q^2 $. The \( \hat{\sigma}(\hat{s}) \) is the partonic cross-section for \( q \bar{q} \to Z' \to \nu_R \nu_R \), with \( \hat{s} = x_1 x_2 s \) is the partonic center-of-mass energy squared, where \( s \) is the total hadronic center-of-mass energy squared. The partonic cross-section is given by:
\begin{equation}
\hat{\sigma}(\hat{s}) = \frac{4 \pi^2}{\hat{s}} \frac{\Gamma(Z' \to q \bar{q}) \Gamma(Z' \to \nu_R \nu_R)}{(M_{Z'}^2 - \hat{s})^2 + M_{Z'}^2 \Gamma_{Z'}^2},
\end{equation}
where \( M_{Z'} \) and \( \Gamma_{Z'} \) are the mass and total decay width of the \( Z' \) boson. The partial decay widths of the \( Z' \) boson
\( \Gamma(Z' \to q \bar{q}) \) and \( \Gamma(Z' \to \nu_R \nu_R) \) are given by
\begin{eqnarray}
\Gamma(Z' \to q \bar{q}) &=& \frac{n_q}{72 \pi} g''^2 M_{Z'},\\
\Gamma(Z' \to \nu_R \nu_R) &=& \frac{n_{\nu_R}}{24 \pi} g''^2 M_{Z'}.
\end{eqnarray}
where \( n_q = 6 \), representing the six active quark flavors, as the \( Z' \) mass is generally greater than twice the top quark mass, and $n_{\nu_R} =3$.

%%%%%%%%%%%%%%%%%%%%%%%%%%%%%%%%%%%%%%%%%%%%%

\section{BDT Method and Signature Analysis}

In this study, we focus on the production of a heavy neutral gauge boson, $Z'$, at the LHC, and its subsequent decay into a pair of RHNs ($\nu_R$). Each $\nu_R$ further decays into a lepton and a $W$ boson, following the chain:
\begin{equation}
	pp \rightarrow Z' \rightarrow \nu_R \nu_R, \quad \nu_R \rightarrow \ell W.
\end{equation}

Depending on the decay modes of the $W$ bosons (hadronic or leptonic), different final states arise. We classify the possible signatures into three categories:
\begin{enumerate}
	\item \textbf{FS1:} $2\ell$ (same/opposite sign) +$4$ jets
	\item \textbf{FS2:} $4\ell +$ MET
	\item \textbf{FS3:} $3\ell + 2$ jets + MET
\end{enumerate}
In our analysis, we restrict leptons to electrons and muons, and consider light jets originating from the hadronization of $u$, $d$, $s$, and $c$ quarks. The missing transverse energy (MET) arises from neutrinos that escape detection, resulting in an imbalance in the transverse momentum.

\begin{figure}[hbt!]
	\centering
	\includegraphics[width=0.3\textwidth]{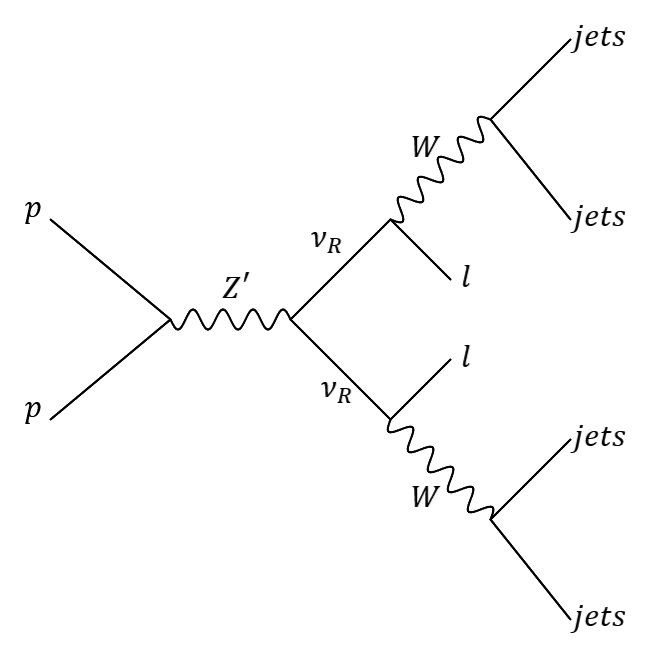}~
	\includegraphics[width=0.3\textwidth]{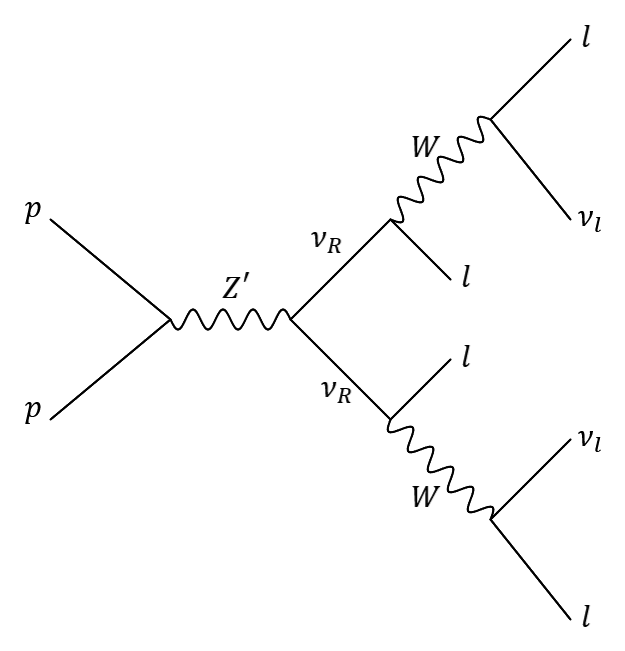}~
	\includegraphics[width=0.3\textwidth]{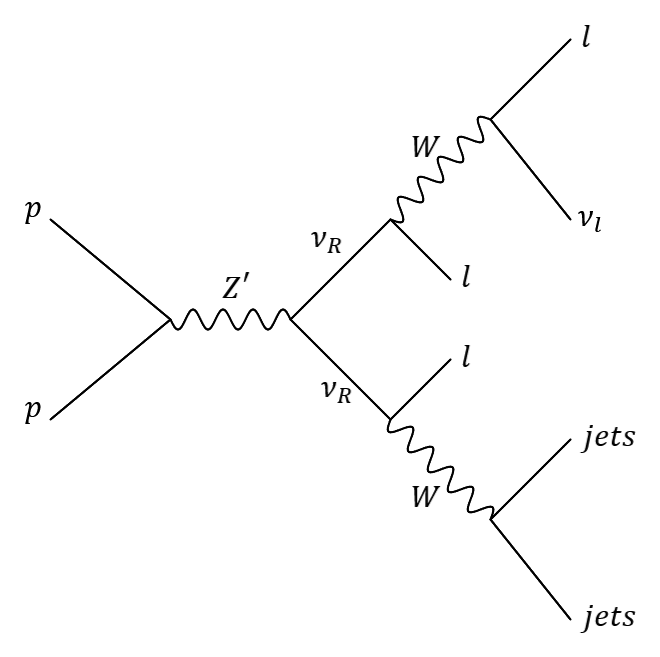}
	\caption{Representative Feynman diagrams of the processes explored in this study.}
	\label{fig:feynman_diagrams}
\end{figure}

Figure~\ref{fig:feynman_diagrams} illustrates the representative Feynman diagrams corresponding to the three final states discussed above.

In the first final state, where both $W$ bosons decay hadronically, we obtain two same-sign or opposite-sign leptons along with four jets. In the remaining two final states, the $W$ bosons decay fully-leptonically and semi-leptonically, respectively. As a result, same-sign charges for whole output leptons are not present, and the final states are characterized by significant MET originating from the undetected neutrinos.

We now discuss the computational setup employed in this study. The BLSM model is implemented in \texttt{SARAH-4.15.1}~\cite{staub2014sarah}, which is used to generate the Feynman rules of the model. The resulting model files are then interfaced with \texttt{SPheno-4.0.5}~\cite{porod2003spheno} to compute the particle mass spectra and full numerical value for model at low energy.

A benchmark point is selected that satisfies current experimental constraints from LHC searches for $Z'$, RHNs ($\nu_R$), additional Higgs bosons, and measurements of the properties of the observed 125\,GeV Higgs boson. The parameters corresponding to the chosen benchmark are listed in Table~\ref{tab:benchmark}.

\begin{table}[hbt!]
	\centering
	\begin{tabular}{|c|c|c|c|}
		\hline
		$M_{Z'}$ & $M_{\nu_{R,1}}$ & $g''\equiv g_{_{B-L}}$ & $\tilde{g}$ \\
		\hline
		3\,TeV & 420\,GeV & 0.42 & $-0.55$ \\
		\hline
	\end{tabular}
	\caption{Parameters of the benchmark point in the BLSM model.}
	\label{tab:benchmark}
\end{table}

To evaluate the LHC sensitivity for this benchmark, we simulate proton-proton collisions at a centre-of-mass energy of 14\,TeV. Event generation for both the signal and background processes is performed using \texttt{MadGraph5\_aMC@NLO}~\cite{alwall2014madgraph5} with its built-in \texttt{Pythia} event generator for radiation, fragmentation, and hadronisation effects, where the BLSM UFO model files are generated via the \texttt{SARAH} framework. Detector effects are simulated using \texttt{Delphes-3.5}~\cite{deFavereau2014delphes}, which reconstructs detector-level objects, including leptons, jets, and MET, forming the observable final states.

In classification problems such as distinguishing BLSM signal events from SM backgrounds, traditional cut-based analyses rely on placing thresholds on individual kinematic variables. However, it is often difficult to identify a single variable with strong discriminating power, and such analyses are limited when the signal-background separation in multi-dimensional phase space is non-rectangular. In these cases, exploiting correlations among multiple variables becomes essential. To address this, we employ the BDT algorithm to analyze the three LHC signatures.

The BDT algorithm extends the cut-based approach by evaluating multiple variables simultaneously and optimizing a corresponding combination for signal-background classification. Instead of discarding events based on a single cut, the BDT recursively selects the most discriminating variable and cut value at each node, splitting events into branches to build a decision tree. This process continues until misclassification is minimized. We provide the BDT with a set of kinematic variables such as the transverse momenta of final-state objects or the invariant mass of lepton pairs as input. The algorithm learns the correlations among these variables for signal and background events and assigns a score between \(-1\) (background-like) and \(+1\) (signal-like). A cut on this BDT score is then used to separate signal from background.

In this study, we employ the \texttt{TMVA} framework~\cite{hoecker2007tmva} within \texttt{ROOT}~\cite{brun1997root} to train a BDT classifier. In the following section, we present a detailed analysis of each of the three final states, including the kinematic variables used for training, their correlations, the model parameters, and the resulting classification performance. In addition, we utilize the \texttt{XGBOOST} algorithm, which simultaneously evaluates multiple kinematic variables to optimize event classification. It constructs decision trees by recursively selecting the most discriminating variables and cut values to separate signal from background, and iteratively refines these using the gradient boosting framework applied to the input feature set.

%%%%%%%%%%%%%%%%%%%%%%%%%%%%%%%%%%%%%%%%%%%%%
\section{$Z'$ and $\nu_R$ signals at the LHC}

In this section, we detail the analysis strategy for the three final states resulting from the BLSM benchmark point considered in this study.

\subsection{FS1: $2\ell + 4j$}

\subsubsection{$2\ell$ (opposite sign) $+ 4j$}

The dominant SM backgrounds for this final state are dileptonic $t\bar{t}$ production with two additional jets ($t\bar{t} + 2j$), and the inclusive $2\ell + 4j$ production. Diboson ($VV + 2j$) and triboson ($VVV$) processes, where $V$ denotes a $W$ or $Z$ boson, can also contribute to this signature. However, in the BLSM signal, the two leptons originate from the decay of a 420\,GeV RHN ($\nu_R$), itself produced in the decay of a heavy $Z'$ boson with mass 3\,TeV. As a result, the leptons carry large transverse momenta, and the dilepton system exhibits a large invariant mass.

We find that imposing the cuts $p_T^{\ell_1}, p_T^{\ell_2} > 200$\,GeV and $M_{\ell\ell} > 250$\,GeV significantly reduces the diboson and triboson backgrounds, while only marginally affecting the signal efficiency. The $t\bar{t} + 2j$ and $2\ell + 4j$ backgrounds exhibit longer tails in these distributions, resulting in residual contributions after these cuts. Also, for completeness, we can consider the background from $W$ + jets, where the $W$ boson produces one genuine lepton and the other lepton arises from a misidentified jet (fake lepton) We have identified such processes in our simulation, in line with what is done experimentally on the CMS and ATLAS results.
 We generate the background samples using \texttt{MadGraph5\_aMC@NLO}, applying the following parton-level cuts:
\begin{equation}
	p_T^{\ell} > 200\,{\rm GeV}, \quad p_T^{j} > 50\,{\rm GeV}, \quad M_{\ell\ell} > 250\,{\rm GeV}.
\end{equation}

Following parton showering and detector simulation, we apply the following baseline event selection criteria:
\begin{eqnarray}
	n_\ell = 2, \quad p_T^{\ell} > 200\,{\rm GeV}, \quad |\eta^\ell| < 2.5, \quad M_{\ell\ell} > 250\,{\rm GeV}, \nonumber \\
	n_j \geq 4, \quad p_T^{j} > 50\,{\rm GeV}, \quad |\eta^j| < 2.5.
\end{eqnarray}
Here, $n_\ell$ and $n_j$ represent the number of selected leptons and jets, respectively.

The cross sections for the signal and background processes used in this analysis are summarised in Table~\ref{tab:cross_sections}.

\begin{table}[hbt!]
	\centering
	\begin{tabular}{|c|c|}
		\hline
		\textbf{Process} & \textbf{Cross section (fb)} \\
		\hline
		Signal ($pp \to Z' \to 2l+4j$) & 0.18 \\
		Background 1 ($t\bar{t} (leptonic) + 2j$) & 23.00 \\
		Background 2 ($2\ell + 4j$) & 0.67 \\
		Background 3 ($W + jets+$ fake lepton) & 782.5 \\
		\hline
	\end{tabular}
	\caption{Cross sections for the signal and background processes considered in the $2\ell $(opposite sign)$+ 4j$ final state analysis.}
	\label{tab:cross_sections}
\end{table}

To enhance the signal-to-background discrimination, we train a Boosted Decision Tree (BDT) classifier and the \texttt{XGBOOST} model using the following input observables:
\begin{itemize}
	\item Number of leptons in the event, $n_{\ell}$
	\item Electric charges of the leptons, $q^\ell$
	\item Transverse momenta of leptons and jets, $p_{T}^{\ell,j}$
	\item Invariant mass of the dilepton system, $M_{\ell\ell}$
	\item Invariant masses of di-jet combinations, $M_{jj}$
	\item Invariant masses of di-jet pairs with either lepton, $M_{\ell jj}$, sensitive to the $\nu_R$ mass
	\item Invariant mass of the complete $2\ell + 4j$ system, $M_{2\ell 4j}$, sensitive to the $Z'$ mass
\end{itemize}

\begin{figure}[hbt!]
	\centering
	\includegraphics[width=0.8\textwidth]{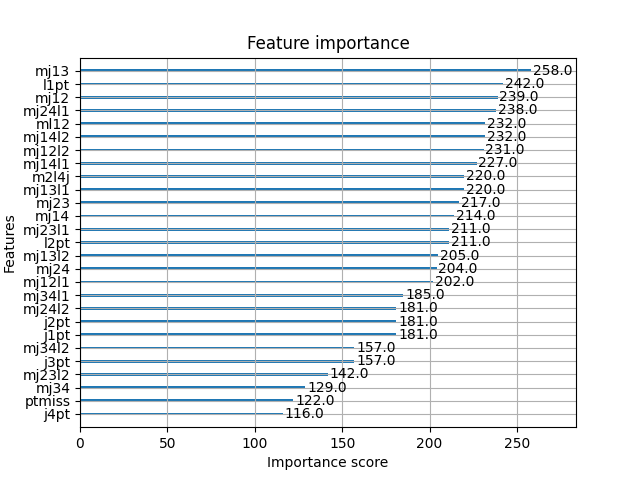}
	\caption{Ranking of the features used in the \texttt{XGBOOST} model training for the BLSM signal for the $2l$(opposite sign)$+4j$ final state against the background.}
	\label{fig:fm_fs1}
\end{figure}

Figure~\ref{fig:fm_fs1} displays the ranking of various features based on their importance in the classification decision, as quantified by the $F$ score—a measure of each variable’s relative significance. Among the most important features, we observe that the invariant masses of the jet pairs, \( m_{j_{1,3}} \) and \( m_{j_{1,2}} \), as well as the transverse momentum of the leading lepton, \( p_T^{\ell_1} \), play the most prominent roles.

\begin{figure}[hbt!]
	\centering
	\includegraphics[width=0.48\textwidth]{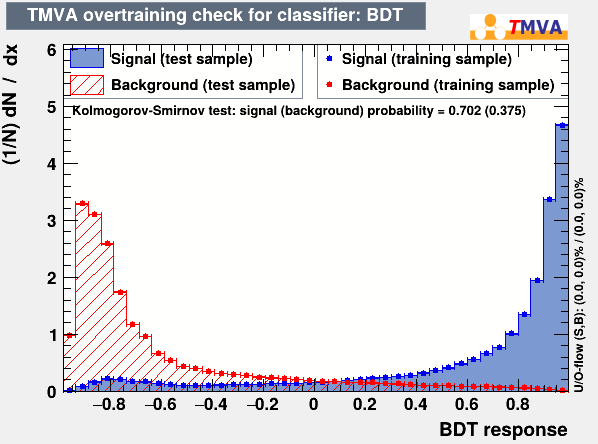}
	\includegraphics[width=0.48\textwidth]{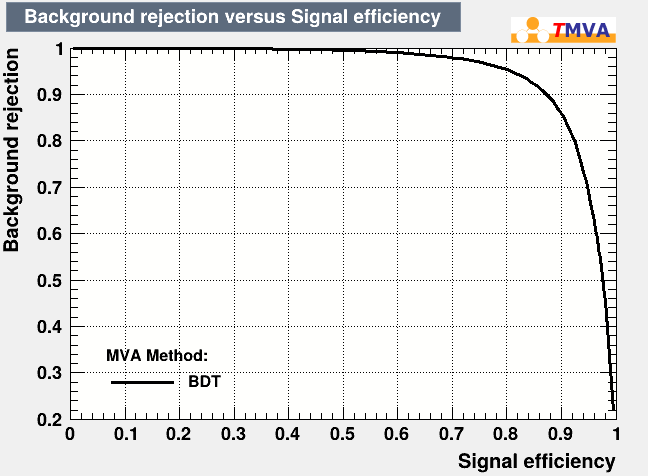}
	\caption{BDT training performance for the $2\ell$ (opposite sign) $+ 4j$ final state. \textbf{Left:} BDT score distribution for signal and background events (training and test samples). \textbf{Right:} ROC curve showing the background rejection versus signal efficiency.}
	\label{fig:BDT_training_FS1}
\end{figure}

In Fig.\,\ref{fig:BDT_training_FS1}, we present the results of the BDT training and testing sample for the signal and background processes. The \textbf{left} panel shows the distribution of the classifier output (BDT score) for signal and background events. We observe that background events predominantly populate the negative side of the distribution, while signal events peak at positive BDT scores, illustrating the separation power of the classifier.

The \textbf{right} panel displays the Receiver Operating Characteristic (ROC) curve, which quantifies the classifier’s performance. For a signal efficiency of 80\%, the BDT achieves a background rejection rate of approximately 95\%. This confirms the effectiveness of the BDT in distinguishing the signal from the background.

\begin{figure}[hbt!]
\centering
\includegraphics[width=0.55\textwidth]{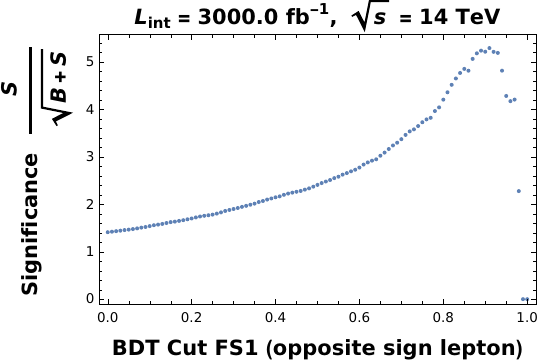}
\caption{Variation of the signal significance at the HL-LHC with BDT threshold on the signal and background events for the final state $2l$ (opposite sign) $+4j$.}
\label{fig:significance_FS1}
\end{figure}

To quantify the expected sensitivity at the HL-LHC, we compute the statistical significance defined as \( S/\sqrt{S+\sum B} \), where \( S \) and \( \sum B \) are the expected number of signal and background events passing a given BDT threshold. In Fig.\,\ref{fig:significance_FS1}, we show the variation of the significance as a function of the BDT cut value, assuming an integrated luminosity of 3000 fb\(^{-1}\).
 
\begin{figure}[hbt!]
	\centering
	\includegraphics[width=0.48\textwidth]{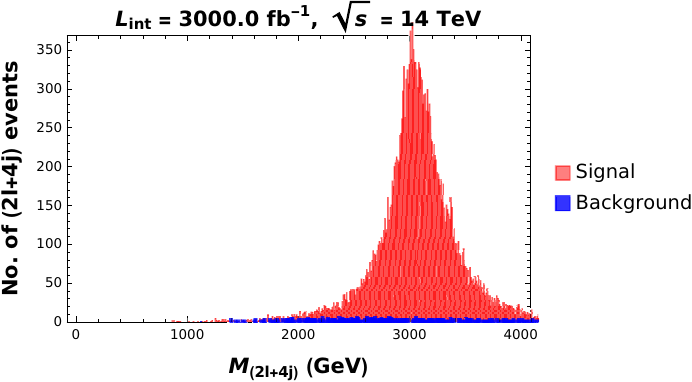}
	\includegraphics[width=0.48\textwidth]{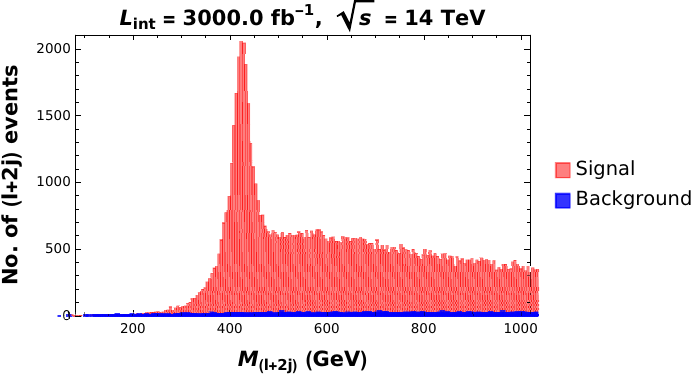}
	\caption{Background and signal distributions for $M_{2l+4j}$ and $M_{2j+l}$ for events having a BDT score greater than 0.6 for the final state $2l$ (opposite sign) $+4j$.}
	\label{fig:mass_dist_FS1}
\end{figure}

Next, we look at the distributions of the variables sensitive to the $Z'$ and $\nu_R$ masses. Fig.\,\ref{fig:mass_dist_FS1} shows the distributions of the invariant masses, $M_{2l+4j}$ and $M_{l+2j}$, for the signal and the background processes with events passing the BDT threshold. We find that the background is suppressed after the BDT cut, and the signal events peak around the $Z'$ mass of 3\,TeV for $M_{2l+4j}$ and the $\nu_R$ mass of around 420\,GeV for $M_{2j+l}$. 

\subsubsection{$2\ell$ (same sign) $+ 4j$}

In this section, we consider the final state consisting of two same-sign leptons and four jets. There are three main SM backgrounds for this final state. 

The first is same-sign $W^\pm W^\pm$ boson pair production in association with two jets ($W^\pm W^\pm jj$), typically produced via Vector Boson Scattering (VBS). 

The second is top quark pair production with an associated $W$ boson ($t\bar{t}W^+$ or $t\bar{t}W^-$), which can yield multiple leptons and jets, including configurations with same-sign leptons. 

The third is the $W$ + jets process, where the $W$ boson decays leptonically to one genuine lepton, and an additional lepton arises from a misidentified jet (fake lepton), resulting in an apparent same-sign lepton pair.

We generate both the signal and background samples using \texttt{MadGraph5}, and apply the same selection cuts used for the $2\ell$ (opposite sign) $+ 4j$ analysis. Additionally, the same set of input observables is used to train the BDT for the same-sign case as in the opposite-sign lepton scenario.

The cross sections for the signal and background processes used in this analysis are summarised in Table~\ref{tab:cross_sectionsoppo}.

\begin{table}[hbt!]
	\centering
	\begin{tabular}{|c|c|}
		\hline
		\textbf{Process} & \textbf{Cross section (fb)} \\
		\hline
		Signal ($pp \to Z' \to 2\ell+4j$) & 0.17 \\
		Background 1 ($W^\pm W^\pm jj$) & 155.7 \\
		Background 2 ($t\bar{t}W^\pm$) & 0.0083 \\
		Background 3 ($W$ + jets + fake lepton) & 782.5 \\
		\hline
	\end{tabular}
	\caption{Cross sections for the signal and background processes considered in the $2\ell$ (same sign) $+ 4j$ final state analysis.}
	\label{tab:cross_sectionsoppo}
\end{table}

\begin{figure}[hbt!]
	\centering
	\includegraphics[width=0.8\textwidth]{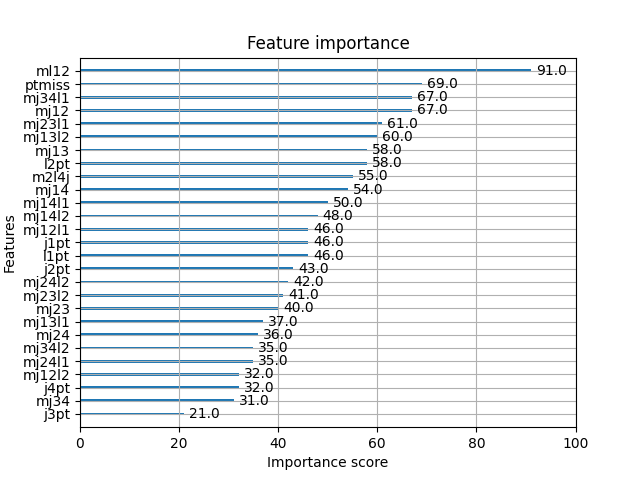}
	\caption{Ranking of the features used in the \texttt{XGBOOST} model training for the BLSM signal for the $2l$(same sign)$+4j$ final state against the background.}
	\label{fig:fm_fs1same}
\end{figure}

Figure~\ref{fig:fm_fs1same} shows the ranking of features by their $F$ score, which indicates their relative importance in the classification for the same-sign lepton case. From the plot, we observe that the most important variables are the invariant mass of the lepton pair, \( m_{\ell_{1,2}} \), and the missing transverse momentum, \( p_T^{\text{miss}} \).

\begin{figure}[hbt!]
	\centering
	\includegraphics[width=0.48\textwidth]{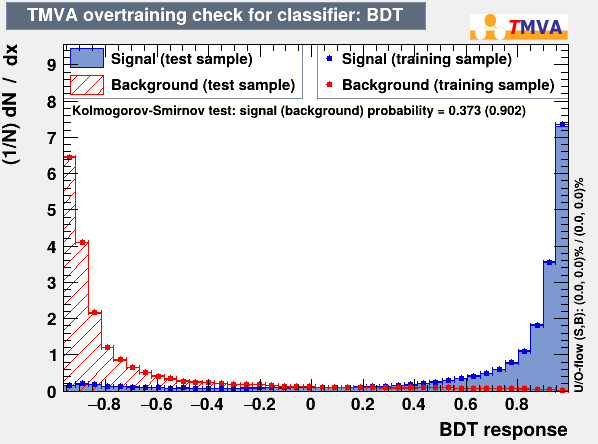}
	\includegraphics[width=0.48\textwidth]{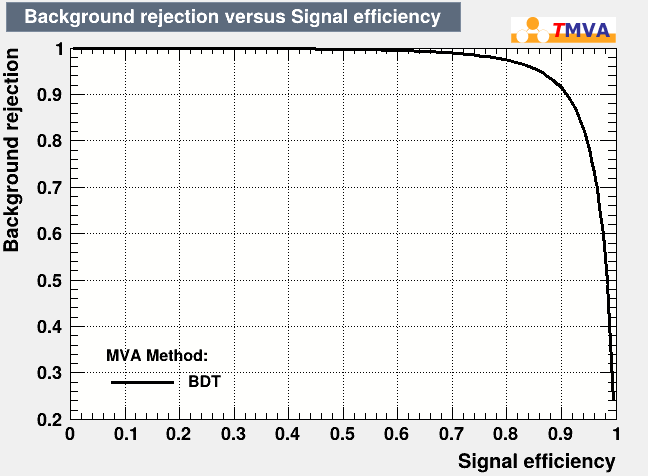}
	\caption{BDT training performance for the $2\ell$ (same sign) $+ 4j$ final state. \textbf{Left:} BDT score distribution for signal and background events (training and test samples). \textbf{Right:} ROC curve showing background rejection versus signal efficiency.}
	\label{fig:BDT_training_FS1same}
\end{figure}

Figure~\ref{fig:BDT_training_FS1same} shows the results of the BDT training and testing samples for both the signal and background processes.

\begin{figure}[hbt!]
	\centering
	\includegraphics[width=0.55\textwidth]{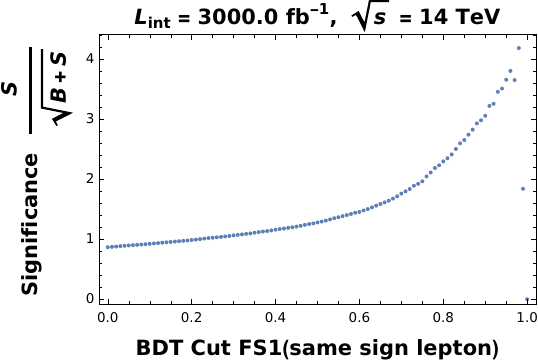}
	\caption{Variation of the signal significance at the HL-LHC with respect to the BDT threshold for the final state $2\ell$ (same sign) $+4j$.}
	\label{fig:significance_FS1same}
\end{figure}

In Figure~\ref{fig:significance_FS1same}, we show the variation of the signal significance as a function of the BDT cut value, assuming an integrated luminosity of 3000 fb\(^{-1}\).

\begin{figure}[hbt!]
	\centering
	\includegraphics[width=0.48\textwidth]{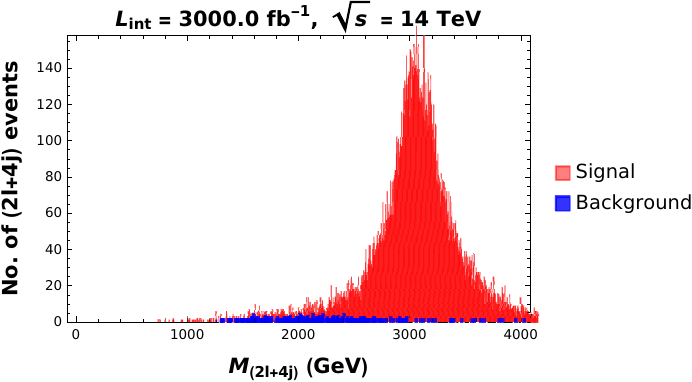}
	\includegraphics[width=0.48\textwidth]{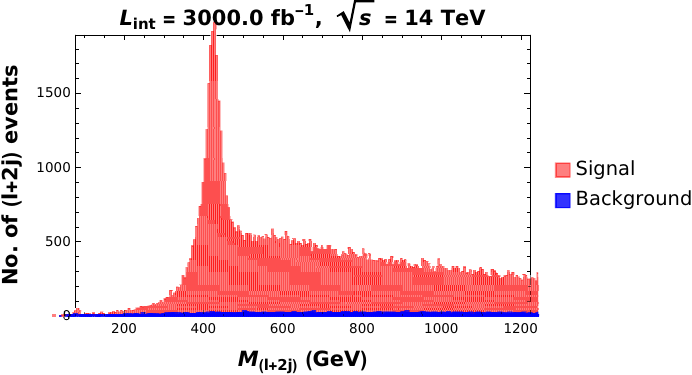}
	\caption{Signal and background distributions for $M_{2\ell+4j}$ and $M_{\ell+2j}$ for events with a BDT score greater than 0.6 in the final state $2\ell$ (same sign) $+4j$.}
	\label{fig:mass_dist_FS1same}
\end{figure}

Finally, in Figure~\ref{fig:mass_dist_FS1same}, we present the invariant mass distributions of $M_{2\ell+4j}$ and $M_{\ell+2j}$ for the signal and background events that pass the BDT selection.

\subsection{FS2: $4\ell +$ MET}

In this final state, we consider SM backgrounds arising from triboson ($VVV$) processes, where $V$ denotes a $W$ or $Z$ boson. Specifically, we include the inclusive $WWZ$, $ZZW$, and $ZZZ$ production processes as backgrounds. In addition, we include the $t\bar{t}Z$ process as an important background, where the $Z$ boson decays leptonically and the top quarks contribute additional leptons and jets.

This channel is particularly clean due to the presence of four high-$p_T$ isolated leptons and MET. Consequently, we begin with a cut-based preselection to reduce background contamination. Since the leptons originating from the decay chain of a single $\nu_R$ tend to be close to each other in the $\eta$-$\phi$ plane, we apply a tight isolation requirement in our \texttt{Delphes} simulation, using a cone radius of 0.1 for lepton isolation.

We impose the following preselection criteria:
\begin{equation}
	n_\ell = 4, \quad p_T^{\ell} > 50\,\text{GeV}, \quad |\eta^{\ell}| < 2.5, \quad p_T^{\text{miss}} > 100\,\text{GeV}.
\end{equation}

The cross sections for the signal and background processes used in this analysis are summarised in Table~\ref{tab:cross_sections2}.

\begin{table}[hbt!]
	\centering
	\begin{tabular}{|c|c|}
		\hline
		\textbf{Process} & \textbf{Cross section (fb)} \\
		\hline
		Signal ($pp \to Z' \to 4\ell + \text{MET}$) & 0.11 \\
		Background 1 ($WWZ$) & 94.2 \\
		Background 2 ($ZZW$) & 30.3 \\
		Background 3 ($ZZZ$) & 10.3 \\
		Background 4 ($t\bar{t}Z$) & 2.15 \\
		\hline
	\end{tabular}
	\caption{Cross sections for the signal and background processes considered in the $4\ell$ final state analysis.}
	\label{tab:cross_sections2}
\end{table}

To further enhance the discrimination between signal and background, we train a Boosted Decision Tree (BDT) classifier and the \texttt{XGBOOST} model using the following set of input observables:

\begin{itemize}
	\item Transverse momentum of each lepton, $p_{T}^{\ell}$
	\item Missing transverse energy, $E_T^{\text{miss}}$
	\item Transverse mass of each lepton combined with MET, $M_T^{\ell}$
	\item Invariant masses of all combinations of lepton pairs, $M_{\ell\ell}$, which are sensitive to the $\nu_R$ mass
	\item Invariant mass of the four-lepton system, $M_{4\ell}$, sensitive to the $Z'$ mass
\end{itemize}

%\ref{fig:fm_fs1} shows the ranking of the various features according to their importance in making the classification decision, where the $F$ score is a measure of the relative importance of the variable. Among the most important features, we have the leading lepton $p_T$, the missing transverse momentum, the dileptonic invariant mass, and one combination of the $2j+l$ invariant mass, $M_{j_1j_4l_1}$.

%\begin{figure}[hbt!]
%	\centering
%	\includegraphics[width=0.8\textwidth]{feature_imp_2l4j}
%	\caption{Ranking of the features used in the \texttt{XGBOOST} model training for the BLSM signal for the $2l+4j$ final state against the $t\bar{t}$(dileptonic) $+2j$ background.}
%	\label{fig:fm_fs1}
%\end{figure}

\begin{figure}[hbt!]
	\centering
	\includegraphics[width=0.8\textwidth]{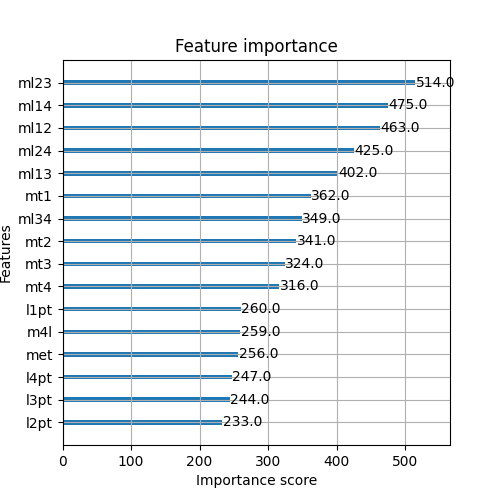}
	\caption{Ranking of the features used in the \texttt{XGBOOST} model training for the BLSM signal for the $4l+$MET final state against the background.}
	\label{fig:fm_fs23}
\end{figure}

Figure~\ref{fig:fm_fs23} shows the ranking of features by their $F$ score, which indicates their relative importance in the classification for the same-sign lepton case. From the plot, we observe that the most important variables are the invariant mass of the lepton pair, \( m_{\ell_{i,j}} \).

\begin{figure}[hbt!]
	\centering
	\includegraphics[width=0.48\textwidth]{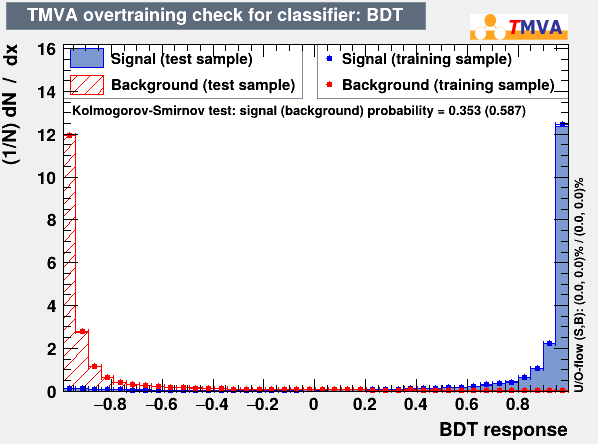}
	\includegraphics[width=0.48\textwidth]{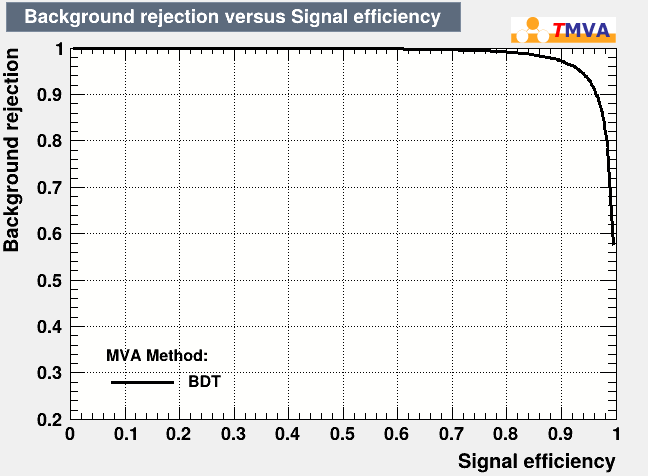}
	\caption{BDT training performance for the $4\ell + MET$ final state. \textbf{Left:} Distribution of the BDT classifier score for signal and background events (both training and testing samples). \textbf{Right:} ROC curve showing background rejection as a function of signal efficiency.}
	\label{fig:BDT_training_FS2}
\end{figure}

In Fig.\,\ref{fig:BDT_training_FS2}, we present the BDT training performance for the $4\ell + MET$ final state (FS2), analogous to the FS1 case discussed previously. The \textit{\bf left} panel shows the BDT score distributions for signal and background events. Compared to FS1, the overlap between the signal and background distributions is smaller, indicating improved discriminating power. This enhanced performance is also reflected in the ROC curve shown in the \textit{\bf right} panel, where we observe that a background rejection of over 95\% can be achieved while maintaining a signal efficiency of about 90\%.

\begin{figure}[hbt!]
	\centering
	\includegraphics[width=0.55\textwidth]{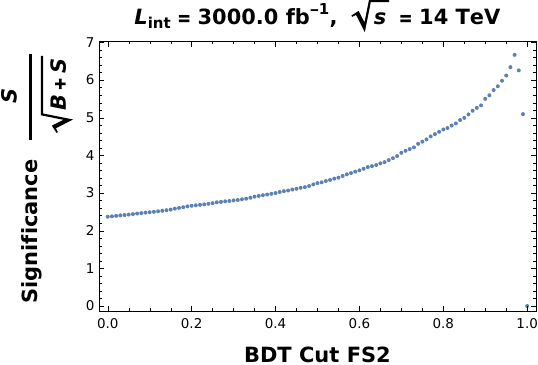}
	\caption{Variation of the expected signal significance at the HL-LHC with the BDT cut value for the $4\ell + MET$ final state.}
	\label{fig:significance_FS2}
\end{figure}

Similarly to the case FS1, and in order to quantify the expected sensitivity at the HL-LHC, we compute the statistical significance using the same formula \( S/\sqrt{S + \sum B} \). The variation of the significance with the BDT cut value is shown in Fig.\,\ref{fig:significance_FS2}.

\begin{figure}[hbt!]
	\centering
	\includegraphics[width=0.48\textwidth]{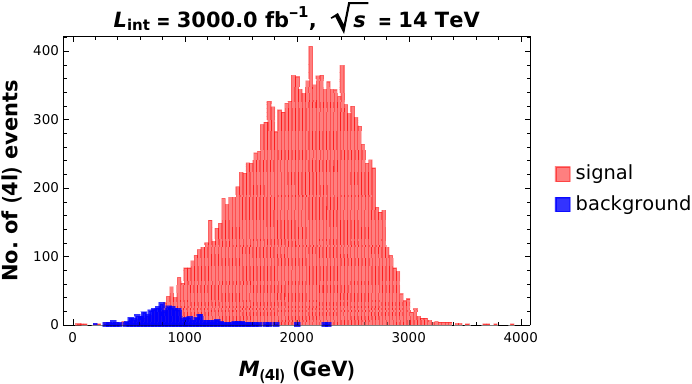}
	\includegraphics[width=0.48\textwidth]{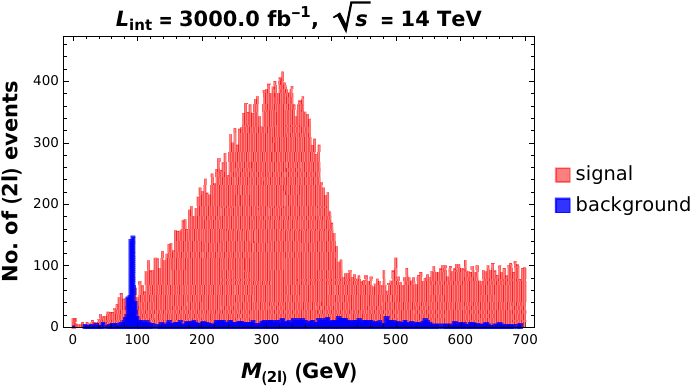}
	\caption{Distributions of $M_{4\ell}$ (left) and $M_{2\ell}$ (right) for signal and background events with BDT score $> 0.6$ in the $4\ell$ final state.}
	\label{fig:mass_dist_FS2}
\end{figure}

In Fig.\,\ref{fig:mass_dist_FS2}, we present the distributions of the invariant masses $M_{4\ell}$ and $M_{2\ell}$ for signal and background events after applying the optimal BDT threshold. We observe a strong suppression of background events due to the BDT cut. Although the presence of neutrinos in the final state leads to significant MET and smears the invariant mass distributions, the signal still exhibits distinctive features: $M_{4 \ell} (M_{2\ell})$, sensitive to $Z' (\nu_R)$ mass, shows a peak around 2 TeV (320 GeV) due to MET instead of 3 TeV (420 GeV). These distributions thus retain valuable information about the underlying resonances, even in the absence of sharp peaks.

\subsection{FS3: $3\ell + 2j + \text{MET}$}

The final state featuring three leptons, two jets, and MET arises when one of the RHNs decays leptonically and the other semi-leptonically. Compared to FS1 and FS2, this signature presents a hybrid topology combining both leptonic and hadronic final states, along with significant MET from the neutrinos.

The dominant SM backgrounds include leptonic $WZ + 2j$ production as well as triboson processes such as $WWZ$, $ZZW$, and $ZZZ$. These channels can mimic the signal when the vector bosons decay in configurations that yield three leptons and two jets in the final state, We also consider the process $t\bar{t}W^\pm$ as background, where the associated $W$ boson decays leptonically, and both top quarks decay via $t \to bW \to b\ell\nu$, leading to a final state with three charged leptons, two $b$-jets, and significant MET.

As in the earlier analyses, the highly boosted nature of the $\nu_R$ due to the heavy $Z'$ (3 TeV) leads to collimated leptons. We apply a minimum separation cut $\Delta R(\ell,\ell) > 0.1$ to account for this. The signal and background events are generated using \texttt{MadGraph5\_aMC@NLO}, applying the following parton-level cuts:
\begin{equation}
	p_T^{\ell} > 50\,\text{GeV}, \quad p_T^j > 50\,\text{GeV}, \quad \Delta R(\ell,\ell) > 0.1.
\end{equation}

Detector effects are included using \texttt{Delphes}, where we further reduce the lepton isolation cone to $\Delta R = 0.1$ to improve lepton reconstruction for boosted decays. After detector simulation, the following event selection criteria are imposed:
\begin{align}
	n_\ell = 3, \quad p_T^\ell > 50\,\text{GeV}, \quad |\eta^\ell| < 2.5,  \nonumber\\
	n_j \geq 2, \quad p_T^j > 50\,\text{GeV}, \quad |\eta^j| < 2.5, \quad p_T^{\text{miss}} > 100\,\text{GeV}. 
\end{align}

The cross sections of the signal and background processes are listed in Table~\ref{tab:cross_sections3}.

\begin{table}[hbt!]
	\centering
	\begin{tabular}{|c|c|}
		\hline
		\textbf{Process} & \textbf{Cross section (fb)} \\
		\hline
		Signal ($pp \to Z' \to 3\ell + 2j + \text{MET}$) & 0.4 \\
		Background 1 ($WZ$ (leptonic) $+ 2j$) & 13.98 \\
		Background 2 ($WWZ$) & 94.2 \\
		Background 3 ($ZZW$) & 30.3 \\
		Background 4 ($ZZZ$) & 10.3 \\
		Background 5 ($t\bar{t}W^\pm$) & 3.9 \\
		\hline
	\end{tabular}
	\caption{Cross sections for the signal and background processes in the $3\ell + 2j + \text{MET}$ final state.}
	\label{tab:cross_sections3}
\end{table}

To distinguish the signal from the background more effectively, we employ a Boosted Decision Tree (BDT) classifier trained and the \texttt{XGBOOST} model on the following set of kinematic variables:

\begin{itemize}
	\item Transverse momentum of leptons and jets, $p_T^{\ell, j}$
	\item Invariant masses of lepton and jet pairs, $M_{\ell\ell}$ and $M_{jj}$
	\item Missing transverse energy, $E_T^{\text{miss}}$
	\item Transverse mass of each lepton with MET, $M_T^\ell$
	\item Invariant mass of the full $3\ell + 2j$ system, $M_{3\ell 2j}$, sensitive to $M_{Z'}$
\end{itemize}

\begin{figure}[hbt!]
	\centering
	\includegraphics[width=0.8\textwidth]{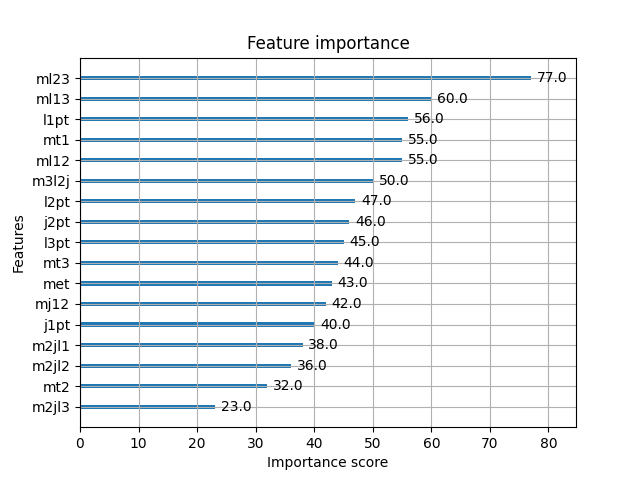}
	\caption{Ranking of the features used in the \texttt{XGBOOST} model training for the BLSM signal for the $3l+2j+$MET final state against the $WZ$(dileptonic) $+2j$ background.}
	\label{fig:fm_fs2}
\end{figure}

Figure~\ref{fig:fm_fs2} shows the ranking of various features based on their importance in the classification decision. From the plot, we observe that the most important variables are the invariant masses of the lepton pair \( m_{\ell_{2,3}} \) and the jet pair \( m_{j_{1,3}} \), along with the transverse momentum of the leading lepton, \( p_T^{\ell_1} \).

\begin{figure}[hbt!]
	\centering
	\includegraphics[width=0.48\textwidth]{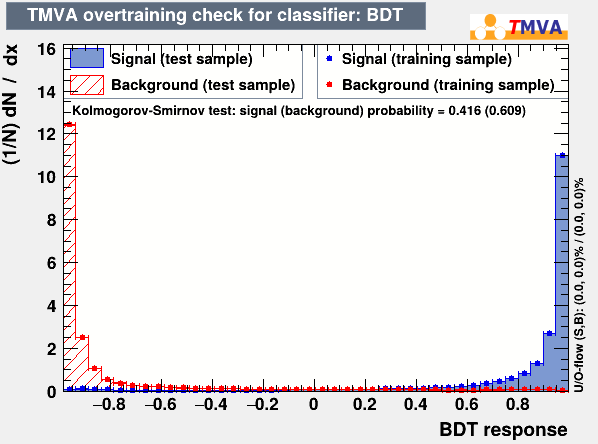}
	\includegraphics[width=0.48\textwidth]{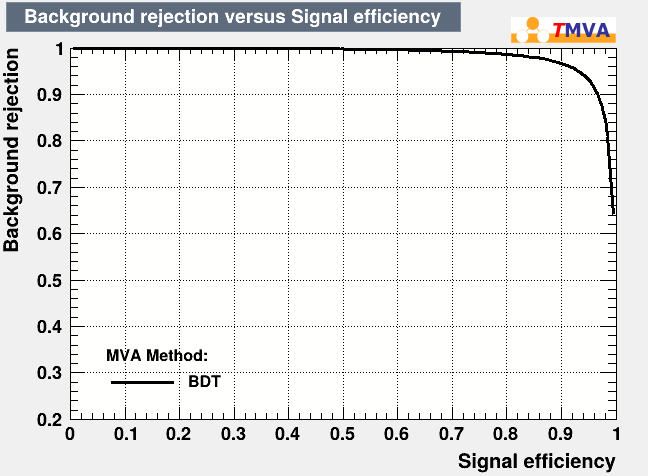}
	\caption{BDT training performance for the $3\ell + 2j + \text{MET}$ final state. \textbf{Left:} BDT score distribution for signal and background events. \textbf{Right:} ROC curve showing background rejection vs. signal efficiency.}
	\label{fig:BDT_training_FS3}
\end{figure}

Figure~\ref{fig:BDT_training_FS3} presents the BDT output for signal and background events. The {\bf left} panel shows the BDT score distribution, where the signal is clearly separated from the background. This separation is more pronounced compared to FS1 and FS2. The {\bf right} panel shows the ROC curve, where for a signal efficiency of 90\%, a background rejection exceeding 98\% is achieved—demonstrating that FS3 provides the strongest discriminative power among the three channels.

\begin{figure}[hbt!]
	\centering
	\includegraphics[width=0.55\textwidth]{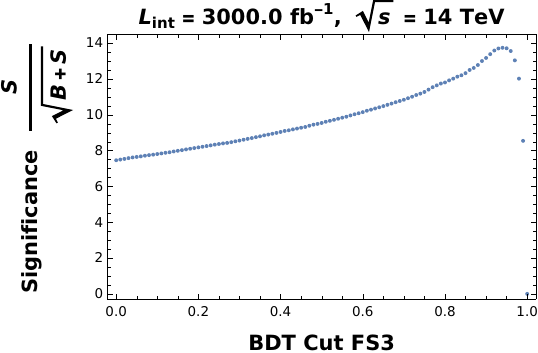}
	\caption{Expected signal significance at the HL-LHC as a function of the BDT score threshold in the $3\ell + 2j + \text{MET}$ channel.}
	\label{fig:significance_FS3}
\end{figure}

The expected signal significance at the HL-LHC is evaluated as a function of the BDT score threshold using the definition \( S/\sqrt{S + \sum B} \). As shown in Fig.~\ref{fig:significance_FS3}, a significance above 3 can be obtained for BDT scores around 0.6, assuming an integrated luminosity of 3000 fb\(^{-1}\).

\begin{figure}[hbt!]
	\centering
	\includegraphics[width=0.48\textwidth]{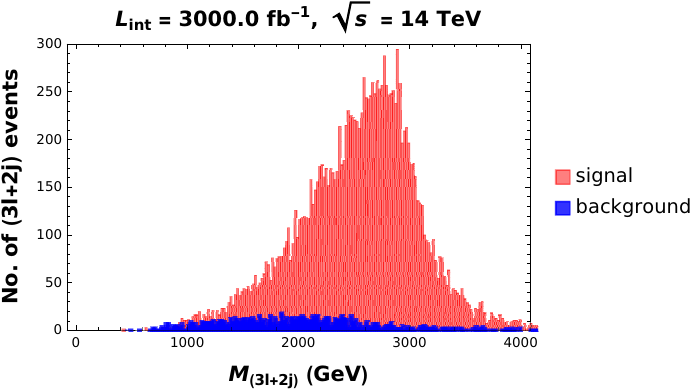}
	\includegraphics[width=0.48\textwidth]{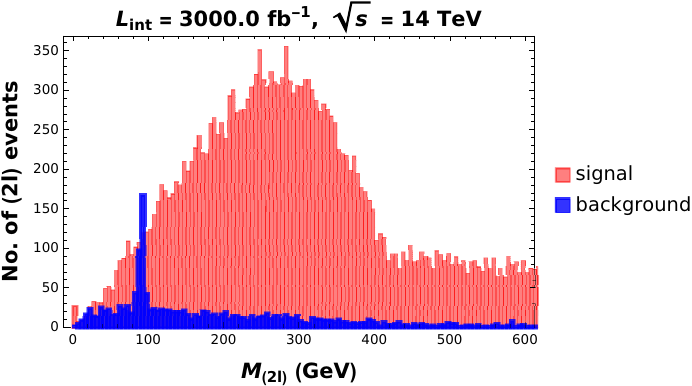}
	\caption{Distributions of $M_{3\ell+2j}$ (left) and $M_{2\ell}$ (right) for events with BDT score $> 0.6$ in the $3\ell + 2j$ final state.}
	\label{fig:mass_dist_FS3}
\end{figure}

Finally, we study the mass-sensitive variables after applying the BDT selection. Figure~\ref{fig:mass_dist_FS3} shows the distributions of $M_{3\ell+2j}$ and $M_{2\ell}$ for the signal and background. $M_{3\ell+2j} (M_{2\ell})$, sensitive to $Z' (\nu_R)$ mass, shows a peak around 2.8 TeV (300 GeV) due to MET instead of 3 TeV (420 GeV), while the backgrounds are heavily suppressed post-BDT selection.

\section{Conclusions}
\label{sec:conclu}

The \( B\mbox{-}L \) extension of the SM (BLSM) offers a natural framework for generating small neutrino masses via the inclusion of RHNs). While experimental searches have constrained the masses and couplings of new particles predicted by the BLSM, such as the heavy \( Z' \) gauge boson and RHNs, significant regions of parameter space remain viable and unexplored. In this study, we examined the discovery potential of a representative BLSM benchmark at the High-Luminosity LHC (HL-LHC), focusing on a region consistent with current experimental bounds. We utilized the \texttt{XGBOOST} framework to assess the discriminating power of various kinematic features and trained a Boosted Decision Tree (BDT) classifier to optimize signal-background separation.

Our analysis considers the production of a \( Z' \) boson in proton-proton collisions, followed by its decay into RHNs, which subsequently decay into leptons and \( W \) bosons. Depending on the \( W \) decay modes, we investigate three final states, combining fully leptonic, semi-leptonic and hadronic outputs. For each channel, we perform a detailed background study and apply a multivariate BDT-based classification strategy to enhance sensitivity. The results indicate that, at the HL-LHC with an integrated luminosity of 3000~fb\(^{-1}\), a signal significance exceeding \( 3\sigma \) can be achieved in each of the considered channels. This highlights the effectiveness of the BDT-based approach in probing BLSM signatures. Our analysis demonstrates that upcoming HL-LHC runs offer promising prospects for discovering new physics predicted by the BLSM. These findings strongly motivate continued experimental efforts to explore this model in future collider searches.

\section*{Acknowledgements}
The work of S. K. and K.E. is partially supported by Science, Technology $\&$ Innovation Funding Authority (STDF) under grant number 48173. N. C.
acknowledges support of the Alexander von Humboldt
Foundation and is grateful for the kind hospitality of the
Bethe Center for Theoretical Physics at the University
of Bonn.

\bibliographystyle{utphys}
\bibliography{ref}

\providecommand{\href}[2]{#2}\begingroup\raggedright\begin{thebibliography}{10}

\bibitem{Fukuda:1998mi}
{\bfseries Super-Kamiokande} Collaboration, Y.~Fukuda {\em et~al.}, ``{Evidence
  for oscillation of atmospheric neutrinos},''
  \href{http://dx.doi.org/10.1103/PhysRevLett.81.1562}{{\em Phys. Rev. Lett.}
  {\bfseries 81} (1998) 1562--1567},
  \href{http://arxiv.org/abs/hep-ex/9807003}{{\ttfamily arXiv:hep-ex/9807003}}.

\bibitem{Ahn:2002up}
{\bfseries K2K} Collaboration, M.~Ahn {\em et~al.}, ``{Indications of neutrino
  oscillation in a 250 km long baseline experiment},''
  \href{http://dx.doi.org/10.1103/PhysRevLett.90.041801}{{\em Phys. Rev. Lett.}
  {\bfseries 90} (2003) 041801}.

\bibitem{Eguchi:2002dm}
{\bfseries KamLAND} Collaboration, K.~Eguchi {\em et~al.}, ``{First results
  from KamLAND: Evidence for reactor anti-neutrino disappearance},''
  \href{http://dx.doi.org/10.1103/PhysRevLett.90.021802}{{\em Phys. Rev. Lett.}
  {\bfseries 90} (2003) 021802}.

\bibitem{Ahmad:2002jz}
{\bfseries SNO} Collaboration, Q.~Ahmad {\em et~al.}, ``{Direct evidence for
  neutrino flavor transformation from neutral current interactions in the
  Sudbury Neutrino Observatory},''
  \href{http://dx.doi.org/10.1103/PhysRevLett.89.011301}{{\em Phys. Rev. Lett.}
  {\bfseries 89} (2002) 011301}.

\bibitem{An:2012eh}
{\bfseries Daya Bay} Collaboration, F.~An {\em et~al.}, ``{Observation of
  electron-antineutrino disappearance at Daya Bay},''
  \href{http://dx.doi.org/10.1103/PhysRevLett.108.171803}{{\em Phys. Rev.
  Lett.} {\bfseries 108} (2012) 171803}.

\bibitem{Mohapatra:1980qe}
R.~N. Mohapatra and R.~E. Marshak, ``{Local B-L Symmetry of Electroweak
  Interactions, Majorana Neutrinos and Neutron Oscillations},''
  \href{http://dx.doi.org/10.1103/PhysRevLett.44.1316}{{\em Phys. Rev. Lett.}
  {\bfseries 44} (1980) 1316--1319}. [Erratum: Phys.Rev.Lett. 44, 1643 (1980)].

\bibitem{Papaefstathiou:2011rc}
A.~Papaefstathiou,
  \href{http://dx.doi.org/10.17863/CAM.16577}{``{Phenomenological aspects of
  new physics at high energy hadron colliders},''} other thesis, 8, 2011.

\bibitem{Masiero:1982fi}
A.~Masiero, J.~F. Nieves, and T.~Yanagida, ``{$B^-$l Violating Proton Decay and
  Late Cosmological Baryon Production},''
  \href{http://dx.doi.org/10.1016/0370-2693(82)90024-7}{{\em Phys. Lett. B}
  {\bfseries 116} (1982) 11--15}.

\bibitem{Mohapatra:1982xz}
R.~N. Mohapatra and G.~Senjanovic, ``{Spontaneous Breaking of Global $B^-$l
  Symmetry and Matter - Antimatter Oscillations in Grand Unified Theories},''
  \href{http://dx.doi.org/10.1103/PhysRevD.27.254}{{\em Phys. Rev. D}
  {\bfseries 27} (1983) 254}.

\bibitem{Buchmuller:1991ce}
W.~Buchmuller, C.~Greub, and P.~Minkowski, ``{Neutrino masses, neutral vector
  bosons and the scale of B-L breaking},''
  \href{http://dx.doi.org/10.1016/0370-2693(91)90952-M}{{\em Phys. Lett. B}
  {\bfseries 267} (1991) 395--399}.

\bibitem{Kang:2015uoc}
Z.~Kang, P.~Ko, and J.~Li, ``{New Avenues to Heavy Right-handed Neutrinos with
  Pair Production at Hadronic Colliders},''
  \href{http://dx.doi.org/10.1103/PhysRevD.93.075037}{{\em Phys. Rev. D}
  {\bfseries 93} no.~7, (2016) 075037},
  \href{http://arxiv.org/abs/1512.08373}{{\ttfamily arXiv:1512.08373
  [hep-ph]}}.

\bibitem{Das:2022rbl}
A.~Das, S.~Mandal, T.~Nomura, and S.~Shil, ``{Heavy Majorana neutrino pair
  production from Z' at hadron and lepton colliders},''
  \href{http://dx.doi.org/10.1103/PhysRevD.105.095031}{{\em Phys. Rev. D}
  {\bfseries 105} no.~9, (2022) 095031},
  \href{http://arxiv.org/abs/2202.13358}{{\ttfamily arXiv:2202.13358
  [hep-ph]}}.

\bibitem{Xia:2018cfz}
L.-G. Xia, ``{Understanding the boosted decision tree methods with the
  weak-learner approximation},''
  \href{http://arxiv.org/abs/1811.04822}{{\ttfamily arXiv:1811.04822
  [physics.data-an]}}.

\bibitem{XIA201915}
L.-G. Xia, ``Qbdt, a new boosting decision tree method with systematical
  uncertainties into training for high energy physics,''
  \href{http://dx.doi.org/https://doi.org/10.1016/j.nima.2019.03.088}{{\em
  Nuclear Instruments and Methods in Physics Research Section A: Accelerators,
  Spectrometers, Detectors and Associated Equipment} {\bfseries 930} (2019)
  15--26}.
  \url{https://www.sciencedirect.com/science/article/pii/S0168900219304309}.

\bibitem{quinlan1986induction}
J.~R. Quinlan, ``Induction of decision trees,'' {\em Machine Learning}
  {\bfseries 1} no.~1, (1986) 81--106.

\bibitem{quinlan1987simplifying}
J.~R. Quinlan, ``Simplifying decision trees,'' {\em International Journal of
  Man-Machine Studies} {\bfseries 27} no.~3, (1987) 221--234.

\bibitem{Basso:2009hf}
L.~Basso, S.~Moretti, and G.~M. Pruna, ``Phenomenology of the minimal b-l
  extension of the standard model: The higgs sector,'' {\em JHEP} {\bfseries
  10} (2009) 006, \href{http://arxiv.org/abs/0903.4777}{{\ttfamily
  arXiv:0903.4777 [hep-ph]}}.

\bibitem{Chun:2017spz}
E.~J. Chun, S.~K. Kang, D.~Y. Kim, and J.~Park, ``Search for $z'$ in type-i
  seesaw models,'' {\em JHEP} {\bfseries 11} (2017) 036,
  \href{http://arxiv.org/abs/1708.00023}{{\ttfamily arXiv:1708.00023
  [hep-ph]}}.

\bibitem{Khalil:2006yi}
S.~Khalil, ``{Low scale $B$ - L extension of the Standard Model at the LHC},''
  \href{http://dx.doi.org/10.1088/0954-3899/35/5/055001}{{\em J. Phys. G}
  {\bfseries 35} (2008) 055001},
  \href{http://arxiv.org/abs/hep-ph/0611205}{{\ttfamily arXiv:hep-ph/0611205}}.

\bibitem{atlas2019zprime}
A.~Collaboration, ``Search for high-mass dilepton resonances using $139$
  $fb^(-1)$ of pp collision data at $13$ tev with the atlas detector,''
  \href{http://dx.doi.org/10.1103/PhysRevLett.123.161801}{{\em Physical Review
  Letters} {\bfseries 123} no.~16, (2019) 161801}.

\bibitem{leike1999zprime}
A.~Leike, ``The phenomenology of extra neutral gauge bosons,''
  \href{http://dx.doi.org/10.1016/S0370-1573(98)00133-1}{{\em Physics Reports}
  {\bfseries 317} (1999) 143--250}.

\bibitem{langacker2009zprime}
P.~Langacker, ``The physics of heavy $z'$ gauge bosons,''
  \href{http://dx.doi.org/10.1103/RevModPhys.81.1199}{{\em Reviews of Modern
  Physics} {\bfseries 81} (2009) 1199--1228}.

\bibitem{cteq1995pdfs}
H.~L. Lai, J.~Botts, J.~Huston, J.~G. Morfin, J.~F. Owens, J.~W. Qiu, W.~K.
  Tung, and H.~Weerts, ``Global qcd analysis and the cteq parton
  distributions,'' \href{http://dx.doi.org/10.1103/PhysRevD.51.4763}{{\em
  Physical Review D} {\bfseries 51} (1995) 4763--4782}.

\bibitem{nnpdf2015pdfs}
R.~D. Ball, V.~Bertone, F.~Cerutti, L.~D. Debbio, S.~Forte, A.~Guffanti,
  J.~Rojo, and M.~Ubiali, ``Parton distributions for the lhc run ii,''
  \href{http://dx.doi.org/10.1007/JHEP04(2015)040}{{\em Journal of High Energy
  Physics} {\bfseries 2015} no.~04, (2015) 040}.

\bibitem{staub2014sarah}
F.~Staub, ``Sarah 4: A tool for (not only susy) model builders,''
  \href{http://dx.doi.org/10.1016/j.cpc.2014.02.018}{{\em Computer Physics
  Communications} {\bfseries 185} (2014) 1773--1790},
  \href{http://arxiv.org/abs/1309.7223}{{\ttfamily arXiv:1309.7223 [hep-ph]}}.

\bibitem{porod2003spheno}
W.~Porod, ``Spheno, a program for calculating supersymmetric spectra, susy
  particle decays and susy particle production at $e^+ e^-$ colliders,''
  \href{http://dx.doi.org/10.1016/S0010-4655(03)00222-4}{{\em Computer Physics
  Communications} {\bfseries 153} (2003) 275--315}.

\bibitem{alwall2014madgraph5}
J.~Alwall, R.~Frederix, S.~Frixione, V.~Hirschi, F.~Maltoni, O.~Mattelaer,
  H.~M.~P. Victoria, T.~Plehn, M.~Schumann, S.~Alioli, and J.~R. Andersen,
  ``The automated computation of tree-level and next-to-leading order
  differential cross sections, and their matching to parton shower
  simulations,'' \href{http://dx.doi.org/10.1007/JHEP07(2014)079}{{\em Journal
  of High Energy Physics} {\bfseries 2014} no.~7, (2014) 79},
  \href{http://arxiv.org/abs/1405.0301}{{\ttfamily arXiv:1405.0301 [hep-ph]}}.

\bibitem{deFavereau2014delphes}
J.~de~Favereau, C.~Delaere, P.~Demin, A.~Giammanco, V.~Lemaître, A.~Mertens,
  and M.~Selvaggi, ``Delphes 3: A modular framework for fast simulation of a
  generic collider experiment,''
  \href{http://dx.doi.org/10.1007/JHEP02(2014)057}{{\em Journal of High Energy
  Physics} {\bfseries 2014} no.~2, (2014) 57},
  \href{http://arxiv.org/abs/1307.6346}{{\ttfamily arXiv:1307.6346 [hep-ex]}}.

\bibitem{hoecker2007tmva}
A.~H{\"o}cker, P.~Speckmayer, J.~Stelzer, A.~Voss, and H.~D. Zech, ``Tmva -
  toolkit for multivariate data analysis,''
  \href{http://dx.doi.org/10.1016/j.physcom.2006.11.005}{{\em Physics
  Communications} {\bfseries 168} (2007) 107--123}.

\bibitem{brun1997root}
R.~Brun and F.~Rademakers, ``Root: An object oriented data analysis
  framework,'' {\em Proceedings of AIHENP'96 Workshop} (1997) 81--86,
  \href{http://arxiv.org/abs/physics/9607061}{{\ttfamily arXiv:physics/9607061
  [physics.data-an]}}.

\end{thebibliography}\endgroup

\end{document}